\documentclass[aps,pra,twocolumn,showpacs,superscriptaddress]{revtex4}
\usepackage{graphicx,dcolumn,bm,amsmath,amssymb,amsfonts}
\begin{document}
\title{
Attainable entanglement of unitary transformed thermal states in
liquid--state nuclear magnetic resonance with the chemical shift 
}
\author{Yukihiro Ota}
\email[Electronic address: ]{oota@hep.phys.waseda.ac.jp}
\affiliation{Department of Physics, Waseda University, Tokyo 169--8555,
Japan} 
\author{Shuji Mikami}
\affiliation{Department of Physics, Waseda University, Tokyo 169--8555,
Japan} 
\author{Motoyuki Yoshida}
\email[Electronic address: ]{motoyuki@hep.phys.waseda.ac.jp}
\affiliation{Department of Physics, Waseda University, Tokyo 169--8555,
Japan} 
\author{Ichiro Ohba}
\email[Electronic address: ]{ohba@waseda.jp}
\affiliation{Department of Physics, Waseda University, Tokyo 169--8555,
Japan}
\affiliation{Kagami Memorial Laboratory for Material Science and
Technology, Waseda University, Tokyo 169--0051, Japan}
\affiliation{Advanced Research Center for Science and Technology, Waseda
University, Tokyo 169--8555, Japan}

\date{\today}
\begin{abstract}
Recently, Yu, Brown, and Chuang [Phys. Rev. A {\bf 71}, 032341 (2005)]
 investigated the entanglement attainable from unitary transformed
 thermal states in liquid--state nuclear magnetic resonance (NMR). 
Their research gave an insight into the role of the entanglement in a
 liquid--state NMR quantum computer. 
Moreover, they attempted to reveal the role of mixed--state entanglement
in quantum computing. 
However, they assumed that the Zeeman energy of each nuclear spin
 which corresponds to a qubit takes a common value for all; there is no
chemical shift. 
In this paper, we research a model with the chemical shifts and
 analytically derive the physical parameter region where unitary
 transformed thermal states are entangled, by the positive partial
 transposition (PPT) criterion with respect to any bipartition. 
We examine the effect of the chemical shifts on the boundary between
the separability and the nonseparability, and find it is negligible. 
\end{abstract}

\pacs{03.67.Mn,\,03.67.Lx}
\maketitle
\section{Introduction}
Quantum mechanics has very different conceptual and mathematical
characters from classical mechanics. 
The superposition and the entanglement (nonseparability) of quantum
states make the difference clear. 
Recently, these characterizations have been utilized for a newly developing
quantum technology.
Actually, quantum entanglement is deeply related to quantum information
processing (QIP)\,\cite{BEZ}. 

The role of entanglement in quantum computing has been researched 
from various viewpoints. 
In particular, the generation of entangled states by quantum dynamics,
or quantum algorithms, will give us useful information on the above
task\,\cite{SSB,YBC}. 
Recently, Yu, Brown, and Chuang\,\cite{YBC} investigated the
entanglement of unitary transformed thermal states in a
liquid--state nuclear magnetic resonance (NMR) quantum computer. 
Such states are defined as the density matrices transformed from the
thermal state in it by a specific class of unitary operators. 
We call the unitary transformations Bell--transformations. 
Their definition will be explained in Sec.\,\ref{subsec:BUTTS}. 
Furthermore, we call such states Bell--transformed thermal states.  
The thermal state in Ref.\,\cite{YBC} is separable (i.e., there is no
quantum correlation)\,\cite{Werner} and characterized by two physical
parameters:   
The one is the number of qubits, and the other is a measure of the
state's polarization. 
The authors in Ref.\,\cite{YBC} studied two kinds of
Bell--transformations.  
One of their central interests is the difference between the
Bell--transformed thermal states and effective pure states.   
The effective pure state is the convex sum of the identity operator and
a pure state, and a typical one is used in the current liquid--state NMR
quantum computer\,\cite{KCL,CGKL,GM}. 
They concluded that the former should be more easily
entangled than the latter; the Bell--transformed thermal states can be
entangled even in the physical parameter region where effective pure
states are separable. 
  
Their research is very important for the following three reasons. 
First, it gives an insight into the role of the entanglement
in a liquid--state NMR quantum computer.  
Braunstein {\it et al}.\,\cite{BCJLPS} pointed 
out that the effective pure states should be almost separable in the current
liquid--state NMR experiments. 
After their study, various studies on the role of entanglement in
liquid--state NMR quantum computing were done (see Ref.\,\cite{YBC}
for additional references). 
One should note that the most natural quantum state is the
thermal one in the liquid--state NMR and an effective pure state is
constructed from it by a sequential operation of several quantum gates. 
Therefore, the authors in Ref.\,\cite{YBC} investigated the entanglement
of a more elementary state than effective pure states in liquid--state NMR
quantum computers. 
Secondly, they attempted to reveal the role of mixed--state entanglement
in quantum computing. 
Its evaluation and the meaning of mixed states will be more subtle than
the case of pure states.
Nevertheless, they can be characterized from the viewpoint of quantum
communication.  
In particular, distillability is
important\,\cite{BBPSSW,HHH1998,ABHHHRWZ}. 
If a mixed state is distillable, it is useful for QIP (e.g.,
quantum teleportation); we can distill a maximal entangled state by
using a number of the copies, and local operations and classical
communication (LOCC). 
Accordingly, their research could lead to the alternative understanding
of mixed--state entanglement from the viewpoint of quantum computing. 
Finally, as mentioned in the beginning of this section, their
work is related to the generation of entanglement by quantum dynamics.  
Actually, they found the difference of the entanglement generation
between the two Bell--transformations.  

In this paper, we analytically derive the physical parameter region where
the Bell--transformed thermal states are entangled in a more general case;
specifically, the effect of the chemical shift\,\cite{GM,VC} is included
in our Hamiltonian.  
The authors in Ref.\,\cite{YBC} assumed that the Zeeman energy of
each nuclear spin which corresponds to a qubit is a common value for
all. 
However, it implies that one can't access the individual qubit; it is
not a realistic model. 
Including such an effect will be necessary to evaluate the
attainability of entanglement from the thermal state in the
experiments. 
Moreover, the analysis taking account of the chemical shift will reveal
how the difference between the Bell--transformations reflect on the
physical parameter region where the Bell--transformed thermal states are
entangled. 
Our method of evaluating the entanglement is simple and straightforward:
Positive partial transposition (PPT) criterion\,\cite{Peres,HHH1996}. 
We analytically calculate the eigenvalues of the partial transposed
Bell--transformed thermal states and find the least ones.  
Here, we would like to emphasize that, in Ref.\,\cite{YBC}, the D\"ur--Cirac
classification\,\cite{DC} has been applied to the evaluation of the
entanglement about two types of Bell--transformed thermal states, but it
doesn't necessarily work in all cases.  

This paper is organized as follows. 
In Sec.\,\ref{sec:Model}, we introduce the Hamiltonian and the
Bell--transformations. 
As in Ref.\,\cite{YBC}, we concentrate on two unitary operators of
Bell--transformations.  
In Sec.\,\ref{sec:nonsep}, in the first, we explain how to specify an
individual bipartition and briefly review the PPT criterion. 
Next, we explain the Bell--diagonal state\,\cite{VPRK,VW2001} in
$N$--qubit system, which plays a central role in this paper. 
Then, we show the main results in Sec.\,\ref{subsec:nonsep}: The
sufficient conditions for the nonseparability of the Bell--transformed
thermal states with respect to any bipartition. 
In Sec.\,\ref{sec:FS_FD}, we show the necessary conditions for the full
separability and the full distillability. 
In particular, we show the necessary and sufficient conditions for their
full distillability when there is no chemical shift. 
In Sec.\,\ref{sec:bd}, we examine the effect of the chemical shift on the
boundary between the separability and the nonseparability determined by
the PPT criterion. 
We summarize our results in Sec.\,\ref{sec:summary}. 
Furthermore, we briefly review the D\"ur--Cirac classification and show
some examples which this method doesn't work in
Appendix.\,\ref{adx:Note_DC}. 

\section{Model}
\label{sec:Model}
\subsection{Hamiltonian}
\label{subsec:Hamiltonian}
In a liquid--state NMR quantum computer, the qubit is the nuclear spin in
the molecule.  
We assume the number of the qubits is $N$ in one molecule.
The dipole--dipole interaction between the molecules in solutions are
negligible because they randomly collide with each other\,\cite{GM}. 
Therefore, we concentrate on the internal degrees of freedom (i.e., the
nuclear spin) of one molecule. 
Let us define the computational bases as $|0\rangle_{i}$ and
$|1\rangle_{i}$ ($\,_{i}\langle 0|0\rangle_{i}=1$, 
$\,_{i}\langle 1|1\rangle_{i}=1$, and 
$\,_{i}\langle 0|1\rangle_{i}=0$). 
The subscription $i(=1,\,2\ldots N)$ is the label of the qubit.
Furthermore, we introduce the following standard operators: 
\(
I_{i}=|0\rangle_{i}\langle 0| + |1\rangle_{i}\langle 1| 
\), 
\(
Z_{i}=|0\rangle_{i}\langle 0| - |1\rangle_{i}\langle 1|
\), 
\(
X_{i}=|0\rangle_{i}\langle 1| + |1\rangle_{i}\langle 0|
\), 
and  
\(
H_{i}=(Z_{i}+X_{i})/\sqrt{2}
\).
We analyze the Hamiltonian
\begin{equation}
\mathcal{H} =  \sum_{i=1}^{N}\frac{h\nu_{i}}{2}Z_{i}, 
\label{eq:Hamiltonian}
\end{equation} 
where $h\nu_{i}$ is the Zeeman energy of the $i$th qubit. 
Note that we neglect the $J$--coupling between the neighboring qubits
because it is much smaller than the Zeeman energy in a liquid--state
NMR\,\cite{GM,VC}.  
In this paper, we don't mention the relationship between the above
computational basis and the physical spin state, and the
Hamiltonian\,(\ref{eq:Hamiltonian}) is a mathematical model. 
We assume $h\nu_{i}$ is positive when we derive our main
results. 
However, the model includes the important physical effect; every
value of $h\nu_{i}$ is different from each other, due to the
chemical shift.
The difference allows us to access each qubit individually.
In Ref.\,\cite{YBC}, every $h\nu_{i}$ is a common value. 
\subsection{Bell--transformed thermal states}
\label{subsec:BUTTS}
Let us consider the separable state which is characterized by a set
of physical parameters. 
Our interest is the parameter region in which the state transformed from
a separable one by quantum gates is a nonseparable one.  
Therefore, we have to specify a suitable initial separable state and
quantum gates which generate the entanglement in a liquid--state NMR
quantum computer. 

First, we refer to the initial separable state. 
The system in a liquid--state NMR experiment is a thermal equilibrium
state with high temperature. 
Accordingly, the most natural choice for the initial state is the
thermal state
\begin{equation}
\rho_{\text{th}}
=
\frac{1}{\mathcal{Z}}e^{-\beta\mathcal{H}}.
\label{eq:thermal}
\end{equation}
In our model, the Hamiltonian $\mathcal{H}$ is given by
Eq.\,(\ref{eq:Hamiltonian}). 
Here, $\beta$ and $\mathcal{Z}=\text{tr}\,e^{-\beta\mathcal{H}}$ are the
inverse temperature and the partition function, respectively. 
Notice that state\,(\ref{eq:thermal}) is a separable state with
respect to any bipartition of the system. 
The physical parameters in Eq.\,(\ref{eq:thermal}) are the number of the
qubit $N$, the Zeeman energy of the $i$th qubit $h\nu_{i}$, and the
inverse temperature $\beta$, which are regarded as the free parameters.
Actually, the values of the parameters are restricted by the experiments for
a liquid--state NMR quantum computing.
For the comparison with Ref.\,\cite{YBC}, we introduce the parameter
$\alpha_{i} = \beta h\nu_{i}/2$, which is called a measure of the
state's polarization in the reference.  
Hereafter, we simply call it polarization. 
The physical meaning of $\alpha_{i}$ is quite clear. 
Let us consider the expectation value of the $z$ component of the total
spin operator $J_{z}(\equiv\sum_{i=1}^{N}Z_{i}/2)$ with respect to
$\rho_{\text{th}}$:  
\(
m_{z}
\equiv
\text{tr}(J_{z}\rho_{\text{th}}) 
= 
-
\sum_{i=1}^{N}(\tanh\alpha_{i})/2
\). 
We assume the value of $N$ is fixed.
The value of $|m_{z}|$ becomes larger as $\alpha_{i}$ increases; the system is
strongly polarized in the direction of the $z$ axis.   
Summarizing the above argument, we can say that the initial thermal
state\,(\ref{eq:thermal}) is characterized by the number of qubits $N$
and the polarization $\alpha_{i}$. 

Next, we explain the unitary operators for generating entanglement. 
We will have to investigate all types of the quantum gates which are
considered as the generator of entanglement and essential parts in
a quantum algorithm.   
However, this task will be very difficult.
In this paper, as in Ref.\,\cite{YBC}, we concentrate on the following
two unitary operators: the controlled--NOT--Hadamard (CH)
transformation, 
$U_{\text{CH}}= U_{\text{fan}} (H_{1}\otimes I)$, 
and the CH--fanout transformation,
$U_{\text{CF}}= U_{\text{CH}}U_{\text{fan}}$. 
The fanout gate $U_{\text{fan}}$ is defined as follows: 
\(
U_{\text{fan}} =
|0\rangle_{1}\langle 0|\otimes I
+
|1\rangle_{1}\langle 1|\otimes X, 
\)
where
\(
I=\bigotimes_{i=2}^{N}\,I_{i}
\),
\(
X=\bigotimes_{i=2}^{N}\,X_{i}
\). 
The above unitary operators are examples in quantum gates,
but they include the essential quantum gates for the generation of
entanglement: controlled--NOT gates and Hadamard gates.
In this case, they generate the entanglement between the first
qubit and the remaining qubits. 
When we consider, for example,  the case of $N=3$ and
that the initial state is 
\(
|000\rangle
=
|0\rangle_{1}\otimes |0\rangle_{2}\otimes |0\rangle_{3}
\), 
we obtain the following results: 
$U_{\text{CH}}|000\rangle=(|000\rangle+|111\rangle)/\sqrt{2}$ and  
$U_{\text{CF}}|000\rangle=(|000\rangle+|111\rangle)/\sqrt{2}$. 

In summary, we examine the entanglement of the states
\begin{eqnarray}
\rho_{\text{CH}} 
&=&
U_{\text{CH}}\rho_{\text{th}}U^{\dagger}_{\text{CH}}, 
\label{eq:rhoCH} \\
\rho_{\text{CF}} 
&=&
U_{\text{CF}}\rho_{\text{th}}U^{\dagger}_{\text{CF}}. 
\label{eq:rhoCF}
\end{eqnarray}
We call the above states the Bell--transformed thermal states. 
The above states are called the Bell--transformed thermal states. 
The initial state $\rho_{\text{th}}$ is separable with any
bipartition of the system for the arbitrary values of $N$ and
$\{\alpha_{i}\}_{i=1}^{N}$.  
\section{Entanglement of Bell--transformed thermal states with chemical shift}
\label{sec:nonsep}
\subsection{Specification of a bipartition}
In order to study the entanglement of a system, it is necessary to
specify the way to divide it into two parts.
We divide the $N$--qubit system into two subsystems, party A and party B, in
the following\,\cite{YBC}.
First, let us consider a set of binary numbers,
$\{k_{i}\}_{i=1}^{N}$ ($k_{i}=0,\,1$).  
When $k_{i}=0$, let the $i$th qubit be in party A. 
On the other hand, when $k_{i}=1$, it is in party B. 
We always set $k_{1}=0$; the first qubit is always in party A. 
For convenience, we introduce an integer
$k=\sum_{i=2}^{N}k_{i}2^{i-2}$.
Therefore, a partition is specified if an integer
$k(\in[1,\,2^{N-1}-1])$ is chosen; we call such a partition the
bipartition $k$.  
Let us choose, for instance, $k=4$ in the case of $N=4$ (i.e.,
$k_{2}=0$, $k_{3}=0$, and $k_{4}=1$). 
The elements of party A are the 1st, 2nd, and 3rd qubit, and 
party B contains only the 4th qubit. 
\subsection{PPT criterion}
\label{subsec:PPT_criterion}
The PPT criterion is the simple and computable way to investigate
entanglement.  
We briefly recapitulate it. 
Let us consider a density matrix $\rho$ in a quantum system with finite
dimension $d$. 
The total system is divided into two subsystems, system A and
system B.
Introducing an orthonormal basis of the system A, 
$\{|u_{i}\rangle_{\text{A}}\}_{i=1}^{d_{\text{A}}}$ and the system B,  
$\{|v_{k}\rangle_{\text{B}}\}_{k=1}^{d_{\text{B}}}$, we can expand the
density matrix $\rho$ as follows: 
\begin{equation}
\rho 
= 
\sum_{i,\,j=1}^{d_{\text{A}}}
\sum_{k,\,l=1}^{d_{\text{B}}}
C(ik|jl)
\,|u_{i}\rangle_{\text{A}}\langle u_{j}| 
\otimes
|v_{k}\rangle_{\text{B}}\langle v_{l}|, 
\label{eq:gen_dm}
\end{equation}
where $C(ik|jl)$ is a complex number and $d=d_{\text{A}}d_{\text{B}}$.
Next, using Eq.\,(\ref{eq:gen_dm}), we define the partial transposition of 
the density matrix with respect to the system B as 
\begin{equation}
\rho^{\text{T}_{\text{B}}}
=
\sum_{i,\,j=1}^{d_{\text{A}}}
\sum_{k,\,l=1}^{d_{\text{B}}}
C(il|jk)
\,|u_{i}\rangle_{\text{A}}\langle u_{j}| 
\otimes
|v_{k}\rangle_{\text{B}}\langle v_{l}|.
\label{eq:PT_genrho}
\end{equation}
Then, we calculate the eigenvalues of $\rho^{\text{T}_{\text{B}}}$ and
investigate their positivity.
If all eigenvalues of Eq.\,(\ref{eq:PT_genrho}) are positive
(i.e., $\rho^{\text{T}_{\text{B}}}\ge 0$), it is called a density matrix
with PPT.   
On the other hand, if at least one eigenvalue of it is negative, it is
called a density matrix with negative partial transposition (NPT). 
The most important thing is that there is the following criterion (i.e.,
PPT criterion): 
\begin{equation}
\rho\text{: separable}\,
\Rightarrow\,
\rho\text{: PPT},
\label{eq:Peres_sep}
\end{equation}
or, equivalently 
\begin{equation}
\rho\text{: NPT}\,
\Rightarrow\,
\rho\text{: entangled (nonseparable)}.
\label{eq:Peres_nonsep} 
\end{equation}
Moreover, the following statement is also known\,\cite{HHH1998,ABHHHRWZ}: 
\begin{equation}
\rho\text{: distillable}\,
\Rightarrow\,
\rho\text{: NPT}. 
\label{eq:Peres_distill}
\end{equation}
\subsection{Bell--diagonal states}
Before showing the our results, we explain the special class of a
density matrix, the Bell--diagonal state\,\cite{VPRK,VW2001}. 
It plays a central role in later discussion. 

To begin, let us explain the generalized
Greenberger--Horne--Zeilinger (GHZ) state\,\cite{YBC,DC} in the
$N$--qubit system  
\begin{equation}
|\Psi^{\pm}_{j}\rangle 
=
\frac{1}{\sqrt{2}}
\left(
|0j\rangle \pm |1\bar{\jmath}\rangle
\right)\quad
(0\le j\le 2^{N-1}-1),
\label{eq:g_GHZ}
\end{equation}
where 
\(
j
=
\sum_{i=2}^{N}j_{i}2^{i-2}
\) 
for the binary number $j_{i}$ ($=0,\,1$), 
\(
|0j\rangle 
=
|0\rangle_{1}\otimes\bigotimes_{i=2}^{N}|j_{i}\rangle_{i}
\) and 
\(
|1\bar{\jmath}\rangle 
=
|1\rangle_{1}\otimes\bigotimes_{i=2}^{N}|1-j_{i}\rangle_{i}
\). 
The symbol $\bar{\jmath}$ means a bit--flip of $j$: $\bar{\jmath}=2^{N-1}-1-j$. 
We can easily find the generalized GHZ states are the elements of an
orthonormal basis of the Hilbert space corresponding to the $N$--qubit system. 

We introduce the following density matrix:
\begin{equation}
\rho_{\text{BD}} 
= 
\sum_{j=0}^{2^{N-1}-1} \left(
\omega_{j}^{+}\lvert\Psi^{+}_{j}\rangle\langle\Psi^{+}_{j}\rvert
+
\omega_{j}^{-}\lvert\Psi^{-}_{j}\rangle\langle\Psi^{-}_{j}\rvert
\right), 
\label{eq:BD_state}
\end{equation}
where 
\(
\omega_{j}^{\pm} = \langle\Psi^{\pm}_{j}|\rho_{\text{BD}}|\Psi^{\pm}_{j}\rangle
\) 
and 
\(
\sum_{j=0}^{2^{N-1}-1}\left(
\omega^{+}_{j}+\omega^{-}_{j}
\right) =1
\). 
Equation (\ref{eq:BD_state}) is a Bell--diagonal state in an $N$--qubit
system. 
We will show that $\rho_{\text{CH}}$ and $\rho_{\text{CF}}$ take the
form of Eq.\,(\ref{eq:BD_state}) in the next subsection. 

The reason why we introduce the Bell--diagonal state is that we can
easily obtain its partial transposition\,\cite{VW2001}.    
We confirm this in the following procedure.
First, let us consider a bipartition $k$.
Secondly, we represent $|\Psi^{\pm}_{j}\rangle\langle\Psi^{\pm}_{j}|$ in the
computational basis: 
\(
|\Psi^{\pm}_{j}\rangle\langle\Psi^{\pm}_{j}|
=
(
|0j\rangle\langle 0j| 
\pm 
|0j\rangle\langle 1\bar{\jmath}|
\pm 
|1\bar{\jmath}\rangle\langle 0j|
+
|1\bar{\jmath}\rangle\langle 1\bar{\jmath}|
)/2
\).
The diagonal parts of $|\Psi^{\pm}_{j}\rangle\langle\Psi^{\pm}_{j}|$ are 
$|0j\rangle\langle 0j|$ and 
$|1\bar{\jmath}\rangle\langle 1\bar{\jmath}|$ 
and are invariant under the partial transposition with respect to party B. 
The off--diagonal ones are 
$|0j\rangle\langle 1\bar{\jmath}|$ and 
$|1\bar{\jmath}\rangle\langle 0j|$ because $j\neq \bar{\jmath}$. 
If the $i$th qubit is in party B (i.e., $k_{i}=1$), the binary
number $j$ of the off--diagonal parts is transformed into
$j_{i}+1(=j_{i}+k_{i})$ modulo $2$ (e.g., $0 + 1 = 1$ and $1+1=0$) by
partial transposition with respect to party B.  
On the other hand, if the $i$th qubit is in the party A (i.e.,
$k_{i}=0$), the corresponding $j_{i}$ is unchanged; 
\(
j_{i} = j_{i} + k_{i}
\). 
As a result, we obtain the following expression: 
\(
(
\lvert\Psi^{\pm}_{j}\rangle\langle\Psi^{\pm}_{j}\rvert
)^{\text{T}_{\text{B}}}
=
(
\lvert\Psi^{+}_{j}\rangle\langle\Psi^{+}_{j}\rvert
+
\lvert\Psi^{-}_{j}\rangle\langle\Psi^{-}_{j}\rvert 
\pm
\lvert\Psi^{+}_{j\oplus k}\rangle\langle\Psi^{+}_{j\oplus k}\rvert
\mp
\lvert\Psi^{-}_{j\oplus k}\rangle\langle\Psi^{-}_{j\oplus k}\rvert
)/2,
\)
where
\(
j\oplus k = \sum_{i=2}^{N} l_{i} 2^{i-2}
\)
($l_{i}\equiv j_{i}+k_{i}\mod 2$). 
Accordingly, the Bell--diagonal state partially transposed with
respect to party B is given by
\begin{eqnarray}
\rho_{\text{BD}}^{\text{T}_{\text{B}}}
&=&
\sum_{j=0}^{2^{N-1}-1}
\left(
\mu^{+}_{j}
\lvert\Psi^{+}_{j}\rangle\langle\Psi^{+}_{j}\rvert
+
\mu^{-}_{j}
\lvert\Psi^{-}_{j}\rangle\langle\Psi^{-}_{j}\rvert
\right), 
\label{eq:PT_BD_state}
\end{eqnarray}
where 
\begin{equation}
\mu^{\pm}_{j} 
=
\frac{\omega_{j}^{+}+\omega_{j}^{-}}{2}
\pm
\frac{\omega_{j\oplus k}^{+}-\omega_{j\oplus k}^{-}}{2}.
\label{eq:eigv_PT_BD_state}
\end{equation}
\subsection{Sufficient conditions for nonseparability}
\label{subsec:nonsep}
First of all, let us show Eqs.\,(\ref{eq:rhoCH}) and (\ref{eq:rhoCF})
are just the Bell--diagonal state.
Using the standard relations 
\(
(H_{1}\otimes I) |0j\rangle
=
(|0j\rangle + |1j\rangle)/\sqrt{2}
\),
\(
(H_{1}\otimes I) |1j\rangle
=
(
|0j\rangle - |1j\rangle)/\sqrt{2}
\), 
\(
U_{\text{fan}}|0j\rangle = |0j\rangle
\), 
and  
\(
U_{\text{fan}}|1j\rangle = |1\bar{\jmath}\rangle
\), 
we obtain the following results: 
\begin{eqnarray}
\langle\Psi^{\pm}_{j}|
\rho_{\text{CH}}
|\Psi^{\pm}_{j^{\prime}}\rangle
&=&
\frac{\delta_{jj^{\prime}}}{\mathcal{Z}}
e^{\mp\alpha_{1}}
e^{-\sum_{i=2}^{N}(-1)^{j_{i}}\alpha_{i}}, 
\label{eq:d_pm_me_CH}
\\
\langle\Psi^{\pm}_{j}|\rho_{\text{CF}}|\Psi^{\pm}_{j^{\prime}}\rangle
&=&
\frac{\delta_{jj^{\prime}}}{\mathcal{Z}}
e^{\mp\alpha_{1}}
e^{\mp\sum_{i=2}^{N}(-1)^{j_{i}}\alpha_{i}}, 
\label{eq:d_pm_me_CF}
\\
\langle\Psi^{\pm}_{j}|
\rho_{\text{CH}}
|\Psi^{\mp}_{j^{\prime}}\rangle
&=&
\langle\Psi^{\pm}_{j}|\rho_{\text{CF}}|\Psi^{\mp}_{j^{\prime}}\rangle
=
0
\label{eq:off_pm_me_CHCF}
, 
\end{eqnarray}
where $0\le j,\,j^{\prime}\le 2^{N-1}-1$.
Consequently, both $\rho_{\text{CH}}$ and $\rho_{\text{CF}}$ are
Bell--diagonal states. 

Let us define the mean values of the polarization of party A, $\xi$
and party B, $\eta$ for given $j$ and $k$ as follows: 
\begin{equation}
\xi = \frac{1}{N-w}\sum_{i\in A_{k}}(-1)^{j_{i}}\alpha_{i},\quad
\eta = \frac{1}{w} \sum_{i\in B_{k}}(-1)^{j_{i}}\alpha_{i}, 
\label{eq:def_xi_eta}
\end{equation}
where \(
A_{k}=\{i\in\mathbb{Z}; k_{i}=0,\,1\le i\le N\}
\), 
\(
B_{k}=\{i\in\mathbb{Z}; k_{i}=1,\,1\le i\le N\}
\). 
In Eq.\,(\ref{eq:def_xi_eta}), we conventionally assign zero to $j_{1}$. 
The number $w$ is the total number of the elements of party B 
($1\le w\le N-1$). 
In other words, it is the hamming weight of $k$ (i.e., the number of
one in $\{k_{i}\}_{i=1}^{N}$). 
Each of $\xi$ and $\eta$ is a function of $j$ if
$N$, $\{\alpha_{i}\}_{i=1}^{N}$, and $k$ are fixed. 

Then, we calculate the eigenvalues of
$\rho_{\text{CH}}^{\text{T}_\text{B}}$ and 
$\rho_{\text{CF}}^{\text{T}_\text{B}}$ with respect to the bipartition $k$. 
Now that we know the general expression\,(\ref{eq:eigv_PT_BD_state}) for
the partially transposed Bell--diagonal state, we easily
obtain the desired results. 
For a given bipartition $k$, the eigenvalue of
$\rho^{\text{T}_{\text{B}}}_{\text{CH}}$ is given by
\begin{equation}
\mu^{\pm}_{\text{CH},\,j} 
=
\frac{1}{\mathcal{Z}}e^{\alpha_{1}}\cosh\alpha_{1}\,e^{-(N-w)\xi}\,
n^{\pm}_{\text{CH}}(\eta),
\label{eq:eig_PT_CH}
\end{equation}
where 
\begin{equation}
n^{\pm}_{\text{CH}}(\eta)
=
e^{-w\eta}
\mp
\tanh\alpha_{1}\,e^{w\eta}.  
\end{equation}
Similary, the eigenvalue of
$\rho^{\text{T}_{\text{B}}}_{\text{CF}}$ is given by
\begin{equation}
\mu^{\pm}_{\text{CF},\,j}
=
\frac{1}{\mathcal{Z}}\,
n^{\pm}_{\text{CF}}(\xi,\,\eta), 
\label{eq:eig_PT_CF}
\end{equation} 
where 
\begin{eqnarray}
n^{\pm}_{\text{CF}}(\xi,\,\eta)
&=&
e^{\mp (N-w)\xi}\cosh(w\eta) \nonumber \\
&&
\qquad
\pm e^{\pm (N-w)\xi}\sinh(w\eta). 
\end{eqnarray}

In general, the relative signs between $\alpha_{i}$s can be different. 
For the latter discussion, we evaluate the range of $\xi$
and $\eta$ for the given $\{\alpha_{i}\}_{i=1}^{N}$ and $k$. 
We rewrite Eq.\,(\ref{eq:def_xi_eta}) as 
\begin{equation*}
\xi = \frac{1}{N-w}\sum_{i\in A_{k}}(-1)^{j_{i}+s_{i}}|\alpha_{i}|,\,\,
\eta = \frac{1}{w}\sum_{i\in B_{k}}(-1)^{j_{i}+s_{i}}|\alpha_{i}|, 
\end{equation*}
where $s_{i}=0$ for positive $\alpha_{i}$ and $s_{i}=1$ for negative
$\alpha_{i}$. 
We readily obtain the inequalities
\begin{equation}
|\xi| \le \frac{1}{N-w}\sum_{i\in A_{k}}|\alpha_{i}|\equiv\xi_{\ast},\quad
|\eta| \le \frac{1}{w}\sum_{i\in B_{k}}|\alpha_{i}|\equiv\eta_{\ast}. 
\end{equation}
The condition for $\xi=\xi_{\ast}(\equiv\xi_{max})$ is easily found:
$j_{i}+s_{i}= 0\mod 2$ for any $i\in A_{k}$. 
Similarly, the conditions for $\eta=\eta_{\ast}(\equiv\eta_{max})$ and
$\eta=-\eta_{\ast}(\equiv\eta_{min})$ are given by 
$j_{i}+s_{i}= 0\mod 2$ and
$j_{i}+s_{i}= 1\mod 2$ for any $i\in B_{k}$, respectively. 
Notice that the minimum value of $\xi$ is not always $-\xi_{\ast}$ because
$j_{1}=0$; it is 
\(
-\xi_{\ast} + [1+(-1)^{s_{1}}]|\alpha_{1}|/(N-w)\equiv \xi_{min}
\).

Now, we analytically derive the sufficient conditions for the
nonseparability of $\rho_{\text{CH}}$ and $\rho_{\text{CF}}$. 
What is needed is that we search for the minimum values
of the eigenvalues which can be negative. 

First, we examine Eq.\,(\ref{eq:eig_PT_CH}). 
In order to discuss definitely, we assume $\alpha_{1}$ is positive for a
while. 
We find that the eigenvalue $\mu^{-}_{\text{CH},\,j}$ is always
positive because all factors of the right hand side are positive. 
Then we focus on $\mu^{+}_{\text{CH},\,j}$. 
The positivity is determined by the value of
$n^{+}_{\text{CH}}(\eta)$. 
Notice that it is a monotonic decreasing function of
$\eta$, because 
\(
\partial n^{+}_{\text{CH}}(\eta)/\partial \eta <0
\). 
Therefore, the minimum value of $n^{+}_{\text{CH}}(\eta)$ is given by 
\(
n^{+}_{\text{CH}}(\eta_{max})
=
n^{+}_{\text{CH}}(\eta_{\ast})
=
e^{-w\eta_{\ast}}-\tanh\alpha_{1}\,e^{w\eta_{\ast}}
\). 
We examine the case of negative $\alpha_{1}$ in turn. 
In the case, the value of $n^{-}_{\text{CH}}(\eta)$ is important for the
examination of the positivity of $\rho_{\text{CH}}$. 
We can readily check that the minimum value of
$\eta^{-}_{\text{CH}}(\eta)$ is given by 
\(
n^{-}_{\text{CH}}(\eta_{max})
=
n^{-}_{\text{CH}}(\eta_{\ast})
=
e^{-w\eta_{\ast}}-\tanh|\alpha_{1}|\,e^{w\eta_{\ast}}
\).
Summarizing the above argument, we can say that $\rho_{\text{CH}}$ is
NPT with respect to the bipartition $k$ if and only if 
\begin{equation}
n^{+}_{\text{CH}}(\eta_{\ast})<0\,
\iff\,
e^{-2w\eta_{\ast}} <\tanh|\alpha_{1}|
\label{eq:NPT_CH_wrt_k}.
\end{equation}

Secondly, let us consider Eq.\,(\ref{eq:eig_PT_CF}). 
We concentrate on the behavior of $n^{\pm}_{\text{CF}}(\xi,\,\eta)$
because $\mathcal{Z}>0$. 
First, we investigate $n^{+}_{\text{CF}}(\xi,\,\eta)$.  
Notice that the value is always positive for $\eta \ge 0$ or  
$\xi,\,\eta\le 0$. 
Hereafter, we consider the case of $\xi>0$ and $\eta<0$.
In this case, we find  
\(
\partial n^{+}_{\text{CF}}(\xi,\,\eta)/\partial \xi < 0
\) 
and 
\(
\partial n^{+}_{\text{CF}}(\xi,\,\eta)/\partial \eta > 0
\). 
Therefore, the minimum value of $n^{+}_{\text{CF}}(\xi,\,\eta)$ is given
by 
\(
n^{+}_{\text{CF}}(\xi_{max},\,\eta_{min})
=
n^{+}_{\text{CF}}(\xi_{\ast},\,-\eta_{\ast})
=
e^{-(N-w)\xi_{\ast}}\cosh(w\eta_{\ast}) - e^{(N-w)\xi_{\ast}}\sinh(w\eta_{\ast})
\). 
Next, we consider $n^{-}_{\text{CF}}(\xi,\,\eta)$. 
It should be noted that the following relation is fulfilled: 
\(
n^{+}_{\text{CF}}(\xi,\,\eta) = n^{-}_{\text{CF}}(-\xi,\,-\eta)
\). 
Therefore, we readily obtain the information on
$n^{-}_{\text{CF}}(\xi,\,\eta)$. 
Through the above arguments, the minimum value of
$n^{-}_{\text{CF}}(\xi,\,\eta)$ is given by  
\(
n^{-}_{\text{CF}}(\xi_{min},\,\eta_{max})
=
n^{-}_{\text{CF}}(\xi_{min},\,\eta_{\ast})
=
e^{-(N-w)\xi_{\ast}+[1+(-1)^{s_{1}}]|\alpha_{1}|}\cosh(w\eta_{\ast})
-
e^{(N-w)\xi_{\ast}-[1+(-1)^{s_{1}}]|\alpha_{1}|}\sinh(w\eta_{\ast})
\).
On the other hand, $n^{-}_{\text{CF}}(\xi_{min},\,\eta_{\ast})$ is
clearly greater than or equal to 
$n^{+}_{\text{CF}}(\xi_{\ast},\,-\eta_{\ast})$. 
Consequently, $\rho_{\text{CF}}$ is NPT with respect to
the bipartition $k$ if and only if 
\begin{eqnarray}
&&
n^{+}_{\text{CF}}(\xi_{\ast},\,-\eta_{\ast})<0 
\nonumber \\
&\iff& 
\cosh\left[
(N-w)\xi_{\ast}-w\eta_{\ast}
\right]
<
\sinh\left(
N\bar{\alpha}
\right),
\label{eq:NPT_CF_wrt_k}
\end{eqnarray} 
where 
\(
N\bar{\alpha}
\equiv
\sum_{i=1}^{N}|\alpha_{i}|
=
(N-w)\xi_{\ast}+w\eta_{\ast}
\). 

We summarize the sufficient conditions for the nonseparability of
$\rho_{\text{CH}}$ and $\rho_{\text{CF}}$ with respect to the
bipartition $k$ in Table\,\ref{table:NPT_wrt_k}. 
Note that the information of the bipartition $k$ is included in
$\xi_{\ast}$ and $\eta_{\ast}$ through $A_{k}$ and $B_{k}$. 
The definitions of $A_{k}$ and $B_{k}$ are explained below
Eq.\,(\ref{eq:def_xi_eta}).  
With respect to the bipartition $k$, $\rho_{\text{CH}}$ is an entangled
state if the inequality\,(\ref{eq:NPT_CH_wrt_k}) is fulfilled, and
$\rho_{\text{CF}}$ is an entangled one if the
inequality\,(\ref{eq:NPT_CF_wrt_k}) is fulfilled. 
The sufficient condition for the nonseparability of $\rho_{\text{CH}}$
is given by $\alpha_{1}$ and the mean value of the magnitude of
polarization for party B, $\eta_{\ast}$; it doesn't depend on
$\alpha_{i}$ ($i\in A_{k}$) in party A, except for $\alpha_{1}$. 
On the other hand, the one of $\rho_{\text{CF}}$ is determined by
$\xi_{\ast}$ and $\eta_{\ast}$; it depends on the mean values of
the magnitude of polarization for party A and party B.
\begin{table}
\caption{Necessary and sufficient conditions for NPT with respect to
 the bipartition $k$ for each Bell--transformation, where 
$\xi_{\ast}=\sum_{i\in  A_{k}}|\alpha_{i}|/(N-w)$, 
$\eta_{\ast}=\sum_{i\in B_{k}}|\alpha_{i}|/w$,
 and $\bar{\alpha}=\sum_{i=1}^{N}|\alpha_{i}|/N$. }
\label{table:NPT_wrt_k}
\begin{ruledtabular}
\begin{tabular}{cc}
Bell--transformation & condition for NPT \\ \hline 
CH &  $e^{-2w\eta_{\ast}} <\tanh|\alpha_{1}|$ \\
CH--fanout & $\cosh\left[
(N-w)\xi_{\ast}-w\eta_{\ast}
\right]
<
\sinh\left(
N\bar{\alpha}
\right) $ \\
\end{tabular}
\end{ruledtabular}
\end{table}

It should be noticed that if every $\alpha_{i}$ is a common value, we
can readily check that both sufficient conditions\,(\ref{eq:NPT_CH_wrt_k}) and
(\ref{eq:NPT_CF_wrt_k}) are equivalent to the corresponding results in
Ref.\,\cite{YBC}. 
\subsection{Characterization of the Bell--transformations}
\label{subsec:Charact_BT}
It is important for the deep understanding of quantum algorithm to
characterize the property of quantum dynamics in terms of entanglement.  
Let us consider the ability and mechanism of the Bell--transformations
$U_{\text{CH}}$ and $U_{\text{CF}}$ to generate entanglement from
the thermal state. 

The authors in Ref.\,\cite{YBC} discussed this ability without taking
account of the chemical shift. 
They concluded that $U_{\text{CF}}$ is a more effective
Bell--transformation than $U_{\text{CH}}$, because the parameter region,
in which $\rho_{\text{CF}}$ is entangled with respect to a given
bipartition, is wider than the corresponding one of $\rho_{\text{CH}}$.  
Such a result is quite natural because the number of the controlled--NOT
gates in $U_{\text{CF}}$ is twice as many as in $U_{\text{CH}}$. 
 
In Sec.\,\ref{subsec:nonsep} the analysis of the sufficient conditions
for the nonseparability of the Bell--transformed thermal states with the
chemical shift reveal the difference between $U_{\text{CH}}$ and
$U_{\text{CF}}$.  
We assume that the maximum value of $\eta$, $\eta_{\ast}$ is given with
respect to a bipartition $k$.  
In the case, let us consider the necessary information of party A to
examine the entanglement of the Bell--transformed thermal states by the
PPT criterion. 
As for $\rho_{\text{CH}}$, from Eq.\,(\ref{eq:NPT_CH_wrt_k}), we find
that the value of $\alpha_{1}$ is only needed; the local information of
party A is required.  
On the other hand, we have to know the maximum value of $\xi$,
$\xi_{\ast}$ to investigate whether $\rho_{\text{CF}}$ is 
entangled or not, according to Eq.\,(\ref{eq:NPT_CF_wrt_k}); the global
information of party A is required.
These observations can lead to the understanding how the
Bell--transformations make party A and party B entangled. 

\section{Full separability and full distillability}
\label{sec:FS_FD}
Let us consider a $N$--particle system. 
A state of this system is called fully separable (or, $N$--separable) if
the corresponding density matrix $\rho$ can be written as a convex
combination of direct product states:  
\begin{equation}
\rho 
= 
\sum_{i}p_{i}\,\bigotimes_{j=1}^{N}\,\rho_{i}^{(j)}\quad
\left(
\sum_{i}p_{i}=1,\,p_{i}\ge 0
\right), 
\end{equation}
where $\rho_{i}^{(j)}$ is the density matrix on the partial Hilbert space
corresponding to the $j$th partile\,\cite{DC,Nagata}. 
One can easily check that a density matrix has PPT with respect to
any bipartition of the system if it is fully separable. 
On the other hand, we call $\rho$ fully distillable if it is distillable
with respect to any bipartition (i.e., we can create a maximal entangled
pair between qubits in party A and party B by a number of the
copies and LOCC), according to Ref.\,\cite{YBC}. 
Through statement\,(\ref{eq:Peres_distill}), we can readily show
that a density matrix has NPT with respect to any bipartition, if it is
fully distillable. 

As in Ref.\,\cite{YBC}, we examine the full separability and the full
distillability of $\rho_{\text{CH}}$ and $\rho_{\text{CF}}$. 
We summarize the necessary condition for the separability of
$\rho_{\text{CH}}$ and $\rho_{\text{CF}}$. 
According to Eqs.\,(\ref{eq:NPT_CH_wrt_k}) and (\ref{eq:NPT_CF_wrt_k}),
the necessary conditions for the separability of $\rho_{\text{CH}}$ and
$\rho_{\text{CF}}$, respectively,  with respect to the bipartition $k$
are given by 
\begin{eqnarray}
&&
e^{-2w\eta_{\ast}} \ge \tanh|\alpha_{1}|,
\label{eq:PPT_CH_wrt_k} \\
&&
\cosh\left[
(N-w)\xi_{\ast}-w\eta_{\ast}
\right]
\ge 
\sinh\left(
N\bar{\alpha}
\right).
\label{eq:PPT_CF_wrt_k}
\end{eqnarray}
Moreover, from Eq.\,(\ref{eq:Peres_distill}), we find that the necessary
conditions for the distillability of $\rho_{\text{CH}}$ and
$\rho_{\text{CF}}$ with respect to the bipartition $k$ are given by
Eqs.\,(\ref{eq:NPT_CH_wrt_k}) and (\ref{eq:NPT_CF_wrt_k}),
respectively. 
Note that the equalities of Eqs.\,(\ref{eq:PPT_CH_wrt_k}) and
(\ref{eq:PPT_CF_wrt_k}) give the boundaries between the separability and
the distillability. 

Let us consider the full separability and the full distillability of
$\rho_{\text{CH}}$.  
The necessary condition for the full separability and the full
distillability is given by 
\(
\min_{k}\left(
e^{-2w\eta_{\ast}}
\right) \ge \tanh|\alpha_{1}|
\) 
and 
\(
\max_{k}\left(
e^{-2w\eta_{\ast}}
\right)
< \tanh|\alpha_{1}|
\), respectively. 
In conclusion, we obtain the necessary conditions for the full
separability and the full distillability of $\rho_{\text{CH}}$ as
follows:  
\begin{eqnarray}
&&
\rho_{\text{CH}}\text{: fully separable}
\Rightarrow
e^{-2b_{max}}\ge \tanh|\alpha_{1}|
\label{eq:CH_FS}
, \\
&&
\rho_{\text{CH}}\text{: fully distillable} 
\Rightarrow
e^{-2b_{min}}
<\tanh|\alpha_{1}|,
\label{eq:CH_FD} 
\end{eqnarray}
where 
\(
b_{max} = \max_{k}(w\eta_{\ast}) =N\bar{\alpha}-|\alpha_{1}|
\) 
and 
\(
b_{min} = \min_{k}(w\eta_{\ast}) = \min_{i\neq 1}(|\alpha_{i}|)
\). 
The value of $w\eta_{\ast}$ takes the maximum value
$b_{max}$ when $k=2^{N-1}-1$ (i.e., $w=N-1$).
On the other hand, it takes the minimum value 
$b_{min}$ when $k_{i}$ corresponding to the minimum value of
$\alpha_{i}$s is 1 and the remainders are 0 (i.e., $w=1$).   

Similarly, we obtain the following results for $\rho_{\text{CF}}$:
\begin{eqnarray}
\!\!\!\!
\rho_{\text{CF}}\text{: fully separable} 
\!
&\Rightarrow&
\!
\cosh 
d_{min}
\ge \sinh\left(
N\bar{\alpha}
\right),
\label{eq:CF_FS}
\\
\!\!\!\!
\rho_{\text{CF}}\text{: fully distillable} 
\!
&\Rightarrow&
\!
\cosh 
d_{max}
<\sinh\left(
N\bar{\alpha}
\right),
\label{eq:CF_FD}
\end{eqnarray}
where 
\(
d_{max} 
=
\max_{k}|(N-w)\xi_{\ast}-w\eta_{\ast}|
\)
and 
\(
d_{min} 
=
\min_{k}|(N-w)\xi_{\ast}-w\eta_{\ast}|
\). 
The value of $|(N-w)\xi_{\ast}-w\eta_{\ast}|$ takes the maximum value,
\(d_{max}\), when the difference between $(N-w)\xi_{\ast}$ and
$w\eta_{\ast}$ is the largest. 
When $\alpha_{i}=\alpha(>0)$ for any $i$, the condition for $d_{max}$ is
quite simple. 
The value of $d_{max}$ is $(N-2)\alpha$, where $k=2^{N-1}-1$ (i.e.,
$w=N-1$) or $k=1$ (i.e., $w=N-1$).
On the other hand, it takes the minimum value, $d_{min}$ when
$(N-w)\xi_{\ast}$ is the closest value to $w\eta_{\ast}$. 
If every $\alpha_{i}$ is a common value, the value of $d_{min}$ is $0$,
where $w=\lfloor N/2\rfloor$. 
Here, the symbol $\lfloor x\rfloor$ means the greatest integer that is less than or
equal to $x\in\mathbb{R}$. 

Our results (\ref{eq:CH_FS})--(\ref{eq:CF_FD}) are the necessary 
conditions for the full separability or the full distillability. 
When $\alpha_{i} = \alpha(>0)$ for any $i$, the authors in Ref.\,\cite{YBC}
showed the sufficient condition for the full distillability, through the
statements proved in Ref.\,\cite{DC}. 
In this case, combinating our results\,(\ref{eq:CH_FD}) and
(\ref{eq:CF_FD}) with theirs, we can obtain the following important
results: 
\begin{eqnarray}
\rho_{\text{CH}}\text{: fully distillable} 
\!\!&\iff &\!\!
e^{-2\alpha}<\tanh\alpha, 
\label{eq:CH_FD_IFF}\\
\rho_{\text{CF}}\text{: fully distillable} 
\!\!&\iff &\!\!
\tanh\alpha > e^{-2(N-1)\alpha}, 
\label{eq:CF_FD_IFF}
\end{eqnarray}
(We can find that the necessary condition for the full distillability of
$\rho_{\text{CF}}$ is 
\( 
\cosh[(N-2)\alpha] < \sinh(N\alpha)
\) from Eq.\,(\ref{eq:CF_FD}) and this inequality is equal to the
corresponding expression in Eq.\,(\ref{eq:CF_FD_IFF}) after a short
calculation). 
Therefore, we obtain the complete physical parameter regions in which
$\rho_{\text{CH}}$ and $\rho_{\text{CF}}$ can be useful for QIP.
\section{Boundary between separability and nonseparability}
\label{sec:bd}
The equalities in the inequalities (\ref{eq:PPT_CH_wrt_k}) and
(\ref{eq:PPT_CF_wrt_k}) imply the boundary between the separability and
the nonseparability in term of the PPT criterion. 
We investigate the effect of the chemical shift on such boundaries.
Hereafter, we assume all $\alpha_{i}$s are positive, for the sake of simplicity. 

Through the above argument, we easily obtain the following expression
of the boundary for $\rho_{\text{CH}}$:  
\begin{equation}
e^{-2w\eta_{\ast}} = \tanh\alpha_{1}.
\label{eq:BD_CH}
\end{equation}
Similarly, the boundary for $\rho_{\text{CF}}$ is given by
\begin{equation}
\cosh\left[
(N-w)\xi_{\ast}-w\eta_{\ast}
\right]
=
\sinh\left(
N\bar{\alpha}
\right).
\label{eq:BD_CF}
\end{equation}

To compare the boundaries with the chemical shift to those without it,
we try a toy model for $\alpha_{i}$. 
The polarization $\alpha_{i}$ divides into a reference value
$\alpha(>0)$ and the deviation $\delta\alpha_{i}$ from it:  
\(
\alpha_{i}
= \alpha + \delta\alpha_{i}
\). 
Here, we regard $x_{i}\equiv\delta\alpha_{i}/\alpha$ as a uniform random
variable in $[-\delta,\,\delta]$ ($0\le \delta <1$). 
Therefore, in our toy model, the polarization of the $i$th qubit
is given by $\alpha_{i}=\alpha(1+x_{i})$. 
Furthermore, the value of $\alpha$ is regarded as the mean value of
polarization.  

Let us explain how to calculate the boundaries. 
In the first, we choose the value of $\delta$; actually, $\delta=0.1$,
$0.01$, and $0$. 
Then, a sequence of random numbers in $[-\delta,\,\delta]$ is
generated by the Mersenne Twister\,\cite{MT}.  
Next, we specify a bipartition. 
For the sake of simplicity, we assume the first $N-w$ qubits are in the
party A and the latter $w$ qubits are in party B. 
Finally, we numerically calculate the value of $\alpha$ satisfying with 
Eqs.\,(\ref{eq:BD_CH}) or (\ref{eq:BD_CF}) for a given $N$ by the
bisection method. 
\begin{figure}[htbp]
\centering
\scalebox{0.80}[0.80]{\includegraphics{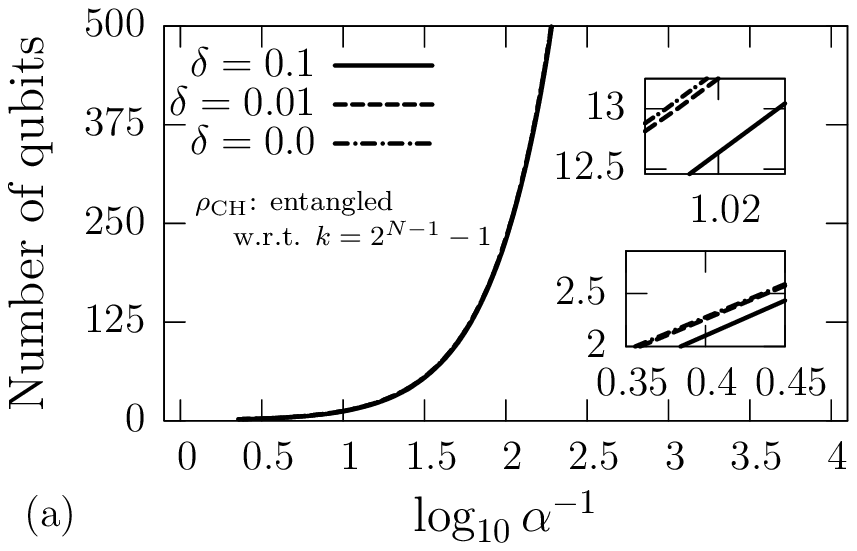}}

\vspace{5mm}
\scalebox{0.80}[0.80]{\includegraphics{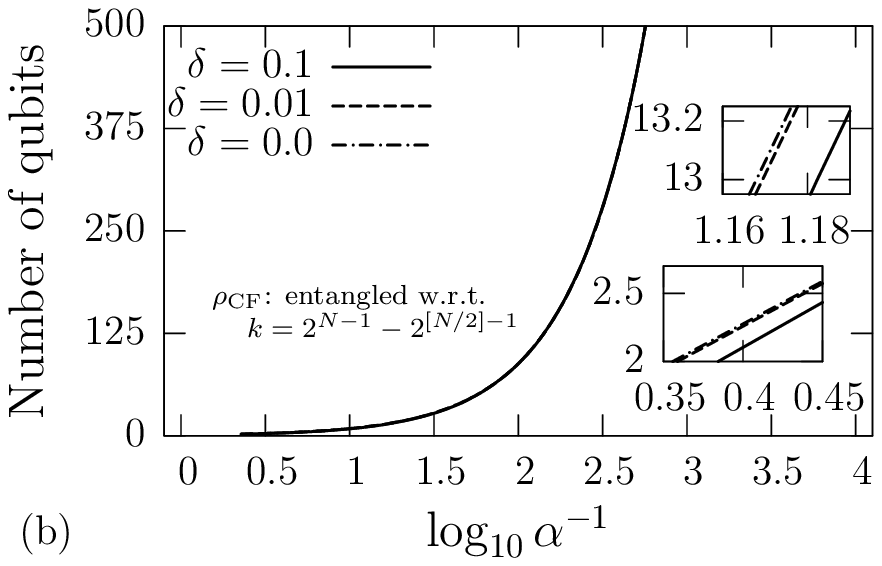}}
\caption{Boundary between the separability and the nonseparability in
 term of the PPT criterion. 
The solid line is for $\delta=0.1$, the broken line is for
 $\delta=0.01$, and the broken dotted line is for $\delta=0$; actually,
 the difference between these is very small (see the insets).  
(a) The CH transformation ($k=2^{N-1}-1$). (b) The CH--fanout transformation
 ($k=2^{N-1}-2^{\lfloor N/2\rfloor-1}$). }  
\label{fig:bound_sep_nonsep}
\end{figure}

In Figs.\,\ref{fig:bound_sep_nonsep}(a) and (b), we show the boundaries on
$(\alpha,\,N)$ plan for a specific sequence of $x_{i}$.
The horizontal axis is the common logarithm of the inverse of $\alpha$.
The larger value of $\log_{10}\alpha^{-1}$ corresponds to the case of
the higher temperature.
The longitudinal axis is the number of qubits. 
The solid line, the broken line, and the broken dotted line correspond
to the boundaries for $\delta=0.1$, $\delta=0.01$, and $\delta=0$,
respectively. 
We show the results for $\rho_{\text{CH}}$ with respect to a bipartition
$k=2^{N-1}-1$ (i.e., $w=N-1$) in Fig.\,\ref{fig:bound_sep_nonsep}(a),
and the case of $\rho_{\text{CF}}$ with respect to a bipartition
$k=2^{N-1}-2^{\lfloor N/2\rfloor-1}$ (i.e., $w=\lfloor N/2\rfloor$) in
Fig.\,\ref{fig:bound_sep_nonsep}(b).   
The left sides of those lines are the parameter regions where
$\rho_{\text{CH}}$ and $\rho_{\text{CF}}$ are nonseparable with respect
to the corresponding bipartition. 
We find that the clear distinction among the different values of
$\delta$ is invisible in both Figs.\,\ref{fig:bound_sep_nonsep} (a) and
(b); for example, denoting the value of $\alpha$ on the boundary as
$\alpha_{b}(\delta)$ for a given $\delta$, we can find that 
\(
|\log_{10}\left[
\alpha_{b}(0)/\alpha_{b}(\delta)
\right]| \le 10^{-2}
\) for $\delta=0.1$ and $0.01$,  
as $N$ is large.
In addition, we can find a similar behavior even if we change the sequence of
random variables and the kind of bipartitions. 
Consequently, the effect of the chemical shift on the boundary between
the separability and the nonseparability is negligible in our model
for $\{\alpha_{i}\}_{i=1}^{N}$. 
This result implies that one have only to examine the number of qubits
$N$ and the mean value of polarization $\alpha$ for the determination of
the entanglement of the Bell--transformed thermal states (\ref{eq:rhoCH})
and (\ref{eq:rhoCF}). 

Let us consider the reason why the effect of the chemical shift is
negligible. 
The boundaries for both $\rho_{\text{CH}}$ and $\rho_{\text{CF}}$
between the separability and the nonseparability are given by
Eqs.\,(\ref{eq:BD_CH}) and (\ref{eq:BD_CF}), respectively. 
Those equations are mainly determined by the mean values of the
polarization of party A, $\xi_{\ast}$ and party B,
$\eta_{\ast}$.  
Therefore, the deviation from the mean value of $\alpha_{i}$s is not
important for the determination of the boundaries. 
However, the model for $\alpha_{i}$ is quite simple; the distribution
of $\alpha_{i}$ is uniform and random. 
We have also assumed that $\alpha_{i}$s are positive in this section. 
Therefore, it is necessary to investigate a more general and realistic
model for $\alpha_{i}$.
\section{Summary}
\label{sec:summary}
We have analytically derived the physical parameter region where the
Bell--transformed thermal states are entangled in the presence of the
chemical shift, by the use of the PPT criterion with respect to any
bipartition.  
Two kinds of Bell--transformations, the CH transformation
$U_{\text{CH}}$ and the CH--fanout transformation $U_{\text{CF}}$, have
been examined, as in Ref.\,\cite{YBC}. 
With respect to the bipartition $k$, $\rho_{\text{CH}}$ is an entangled
state if the inequality\,(\ref{eq:NPT_CH_wrt_k}) is fulfilled, and
$\rho_{\text{CF}}$ is an entangled one if the
inequality\,(\ref{eq:NPT_CF_wrt_k}) is fulfilled. 
We summarize our results in Table \ref{table:NPT_wrt_k}. 
If the every $\alpha_{i}$ is a common value, our results are
equal to the corresponding ones in Ref.\,\cite{YBC}. 
There exists an obvious differences between Eqs.\,(\ref{eq:NPT_CH_wrt_k})
and (\ref{eq:NPT_CF_wrt_k}) with respect to their dependence on $\alpha_{i}$.   
The sufficient condition for the nonseparability of $\rho_{\text{CH}}$
is given by $\alpha_{1}$ and the mean value of $|\alpha_{i}|$ for party
B, $\eta_{\ast}$; it doesn't depend on $\alpha_{i}$ ($i\in A_{k}$) 
in party A, except for $\alpha_{1}$.   
On the other hand, the sufficient condition of $\rho_{\text{CF}}$ is
determined by $\xi_{\ast}$ and $\eta_{\ast}$, the mean values of
$|\alpha_{i}|$ for party A and party B, respectively.

Next, we have derived the necessary conditions for the full 
separability and the full distillability through the above analytical
results. 
On the other hand, the authors in Ref.\,\cite{YBC} obtained the
sufficient conditions when all of $\alpha_{i}$s are equal. 
We have obtained the necessary and sufficient conditions
for the full distillability of $\rho_{\text{CH}}$ and $\rho_{\text{CF}}$
in this case, combinating our results with theirs. 
Accordingly, we can completely characterize the physical parameter
regions in which those unitary transformed thermal states can be useful
for QIP, if there is no chemical shift.
When the chemical shift exists, we will have to examine the sufficiency
of our results. 

Finally, we have investigated the effect of the chemical shift on the
boundary between the separability and the nonseparability determined by
the PPT criterion. 
We have shown such an effect on the boundaries should be negligible. 
Actually, one have only to know the number of qubits $N$ and the mean
value of the magnitude of polarization $|\alpha_{i}|$ for the evaluation
of the entanglement of the Bell--transformed thermal states
(\ref{eq:rhoCH}) and (\ref{eq:rhoCF}). 
This result is quite natural, because the boundaries for both
$\rho_{\text{CH}}$ and $\rho_{\text{CF}}$ between the separability and
the nonseparability dominantly depend on the mean values of
$|\alpha_{i}|$ for party A, $\xi_{\ast}$ and party B, $\eta_{\ast}$.  
We will have to research a more general and realistic model for
$\alpha_{i}$ for examining the effect of the chemical shifts on the
boundary between the separability and the nonseparability. 

It is also important to examine the difference between $\rho_{\text{CH}}$ and
$\rho_{\text{CF}}$ (i.e., $U_{\text{CH}}$ and $U_{\text{CF}}$) in terms of
the entanglement generation. 
The authors in Ref.\,\cite{YBC} concluded that $U_{\text{CF}}$ is a more
effective Bell--transformation than $U_{\text{CH}}$ because the
parameter region in which $\rho_{\text{CF}}$ is fully distillable is
wider than the corresponding one of $\rho_{\text{CH}}$. 
The analysis taking account of the chemical shift has revealed
the difference between $U_{\text{CH}}$ and $U_{\text{CF}}$ from another
point of view.  
When the value of $\eta_{\ast}$ is given with respect to a fixed
bipartition $k$, the value of $\alpha_{1}$ is necessary to examine the
entanglement of $\rho_{\text{CH}}$; the local information of party A is
required. 
On the other hand, for $\rho_{\text{CF}}$, the mean value of
$|\alpha_{i}|$ for party A, $\xi_{\ast}$ is essential; the global
information of party A is required. 
In this paper, we have obtained the analytical expressions of the
eigenvalues of $\rho_{\text{CH}}^{\text{T}_{\text{B}}}$ and
$\rho_{\text{CF}}^{\text{T}_{\text{B}}}$. 
Therefore, the results allow us to characterize $U_{\text{CH}}$ and
$U_{\text{CF}}$ in more detail; for instance, we can evaluate the
negativity, which is an entanglement measure\,\cite{ZHSL,VW}.
We will show the results in the near future. 

Research on entanglement in liquid--state NMR involves various 
aspects of quantum information theory, for example, the role of mixed states in 
quantum computing and the classification of entanglement. 
One should note that the achievable range of the physical parameters $N$
and $\alpha_{i}$s is limited in a current liquid--state NMR experiment. 
Actually, it may be difficult to compare the theoretical results with
the experiments. 
However, several experimental developments have been reported in
liquid--state NMR, for example, the highly polarized initial
states\,\cite{ABCDHJKT} and the number of qubits greater than
ten\,\cite{LK,NMRDCPBHCL}.  
Furthermore, research on a solid--state NMR quantum
computer\,\cite{LGOM,BMRLRHC}, which can relax the limitation of liquid--state
NMR, has been developed steadily. 
Consequently, we expect that theoretical research on the entanglement
in liquid--state NMR could be connected with these experiments in future.
\begin{acknowledgments}
The authors acknowledge H. Nakazato for valuable discussions.  
One of the authors (Y. O.) greatly acknowledges the discussion with
 T. Y. Petrosky, N. Hatano, and M. Machida. 
One of the authors (S. M.) thanks Y. Tanaka. 
This research is partially supported by a Grant--in--Aid for Priority Area B
 (No.\,763), MEXT, and one for the 21st--Century COE Program (Physics of
 Self-Organization Systems) at Waseda University from MEXT, and by a
 Waseda University Grant for Special Research Projects
 (No.\,2004B--872).   
\end{acknowledgments}

\appendix
\section{Notes on D\"ur--Cirac classification}
\label{adx:Note_DC}
The D\"ur--Cirac classification\,\cite{DC} is a very effective way to
evaluate the separability or the distillability of density matrices,
of either pure or mixed states, in a multiqubit system, and it is widely
used\,\cite{YBC,Nagata,BL}. 
However, we have to take care of using it; we don't always obtain information
on their entanglement from the method proposed in Ref.\,\cite{DC}. 
In the appendix, we show several examples in which the method doesn't work. 
Unfortunately, we don't know what kind of density matrices have the
problem.  
Nevertheless, in the end of the appendix, we propose the prescription to
solve this problem in specific examples. 

First of all, we summarize the strategy of the D\"ur--Cirac classification. 
In order to evaluate the entanglement of a quantum state by the PPT
criterion, one must examine the positivity of the partial transposed
density matrices. 
In general, such tasks will be difficult as the number of qubits becomes
large and many choices of bipartitions exist.
On the other hand, if one uses the D\"ur--Cirac classification, it is
only necessary to calculate some specific matrix elements of the state
concerned. 
The main idea is that, using a sequence of local operations, one can
transform an arbitrary density matrix $\rho$ in a $N$--qubit system into
the following state whose property of entanglement is easily examined: 
\begin{eqnarray}
\rho_{N}
&=&
\lambda_{0}^{+}
\lvert\Psi_{0}^{+}\rangle\langle\Psi_{0}^{+}\rvert 
+
\lambda_{0}^{-}
\lvert\Psi_{0}^{-}\rangle\langle\Psi_{0}^{-}\rvert \nonumber \\
&&
\,
+ 
\sum_{j=1}^{2^{N-1}-1}\lambda_{j}
\big(
\lvert\Psi_{j}^{+}\rangle\langle\Psi_{j}^{+}\rvert
+
\lvert\Psi_{j}^{-}\rangle\langle\Psi_{j}^{-}\rvert
\big).
\label{eq:ds} 
\end{eqnarray}
The original density matrix $\rho$ is related to $\rho_{N}$ by the
following expressions: 
\(
\lambda^{\pm}_{0}
=
\langle\Psi^{\pm}_{0}|\rho|\Psi^{\pm}_{0}\rangle
\) and 
\(
2\lambda_{j}
=
\langle\Psi^{+}_{j}|\rho|\Psi^{+}_{j}\rangle 
+
\langle\Psi^{-}_{j}|\rho|\Psi^{-}_{j}\rangle
\). 
The property of $\rho_{N}$ is characterized as follows:
\begin{eqnarray}
\rho_{N}\text{: PPT with a bipartition } k 
\iff
\Delta \le 2\lambda_{k}, 
\label{eq:dsPPT}
\end{eqnarray} 
or
\begin{eqnarray}
\rho_{N}\text{: NPT with a bipartition }k 
\iff
\Delta > 2\lambda_{k},
\label{eq:dsNPT} 
\end{eqnarray} 
where $\Delta=|\lambda^{+}_{0}-\lambda^{-}_{0}|$.
The most important key idea is that the entanglement does not increase
through local operations.  
Accordingly, if $\rho_{N}$ is a nonseparable state with a bipartition,
$\rho$ is also such a state.  
It should be noticed that we obtain information on the entanglement
of $\rho$ only if $\rho_{N}$ is a nonseparable state.

Now, we show three examples. 
The first one is the case in which the D\"ur--Cirac classification works
well. 
The remaining two are not such cases.
Hereafter, we concentrate on a two--qubit system. 
Therefore, the value of $k$ in Eqs.\,(\ref{eq:dsPPT}) and (\ref{eq:dsNPT})
is always $1$. 
In the first, let us consider the following state: 
\begin{equation}
\rho_{iso} 
= 
(1-f)\frac{1}{4}(I_{1}\otimes I_{2}) + f|\Psi^{+}_{0}\rangle\langle\Psi^{+}_{0}|,
\end{equation}
where $-1/3\le f\le 1$. 
The above density matrix is called an isotropic state\,\cite{ABHHHRWZ}.
Directly using the PPT criterion\,\cite{Peres,HHH1996}, we readily find
$\rho_{iso}$ is an entangled state if $f>1/3$.
On the other hand, we apply the D\"ur--Cirac method to $\rho_{iso}$.  
We obtain the following results:  
\(
\langle\Psi^{+}_{0}|\rho_{iso}|\Psi^{+}_{0}\rangle 
= 
(1+3f)/4
\), 
\(
\langle\Psi^{-}_{0}|\rho_{iso}|\Psi^{-}_{0}\rangle 
= 
(1-f)/4
\), 
and 
\(
\langle\Psi^{+}_{1}|\rho_{iso}|\Psi^{+}_{1}\rangle 
+
\langle\Psi^{-}_{1}|\rho_{iso}|\Psi^{-}_{1}\rangle 
=
(1-f)/2
\).
Accordingly, using Eq.\,(\ref{eq:dsNPT}), we also find $\rho_{iso}$
is an entangled one if $f>1/3$. 

Next, we consider a slight different state from $\rho_{iso}$
\begin{equation}
\rho^{\prime}_{iso}
=
(1-f)\frac{1}{4}(I_{1}\otimes I_{2}) + f|\Psi^{+}_{1}\rangle\langle\Psi^{+}_{1}|,
\end{equation} 
where $-1/3 \le f\le 1$. 
Notice that the condition for the nonseparability of
$\rho_{iso}^{\prime}$ is the same one as $\rho_{iso}$; 
$\rho^{\prime}_{iso}$ is an entangled state if $f>1/3$. 
This result is quite natural, because the state $\rho_{iso}$ is transformed
into $\rho^{\prime}_{iso}$ by a local unitary operator; 
\(
\rho^{\prime}_{iso} 
= 
(I_{1}\otimes X_{2})
\rho_{iso}
(I_{1}\otimes X_{2})^{\dagger}
\).
On the other hand, we obtain the eigenvalues in the form of
Eq.\,(\ref{eq:ds}) as follows: 
\(
\langle\Psi^{+}_{0}|\rho^{\prime}_{iso}|\Psi^{+}_{0}\rangle 
= 
(1-f)/4
\), 
\(
\langle\Psi^{-}_{0}|\rho^{\prime}_{iso}|\Psi^{-}_{0}\rangle 
= 
(1-f)/4
\), 
and 
\(
\langle\Psi^{+}_{1}|\rho^{\prime}_{iso}|\Psi^{+}_{1}\rangle 
+
\langle\Psi^{-}_{1}|\rho^{\prime}_{iso}|\Psi^{-}_{1}\rangle 
=
(1+f)/2
\).
Therefore, the value of $\Delta$ is less than $2\lambda_{1}$ for the
arbitrary value of $f$ and the condition (\ref{eq:dsPPT}) is satisfied. 
This might imply that $\rho^{\prime}_{iso}$ is
always separable.
However, $\rho^{\prime}_{iso}$ is local unitary equivalent to the
entangled state $\rho_{iso}$ ($f>1/3$). 
In conclusion, we can't obtain information on the entanglement of
$\rho^{\prime}_{iso}$ by the method in Ref.\,\cite{DC}, because
the entangled state $\rho^{\prime}_{iso}$ is transformed into a
separable state by LOCC. 

The third example is related to the task in this paper. 
We consider the case of $N=2$ and $\alpha_{1}=\alpha_{2}=\alpha$ in
Eq.\,(\ref{eq:rhoCH}). 
We have known the condition for the nonseparability of
$\rho_{\text{CH}}$; $\rho_{\text{CH}}$ is an entangled state if 
\(
e^{-2\alpha}>\tanh\alpha
\).
However, the value of $\Delta$ is less than $2\lambda_{1}$ for any
$\alpha$ because 
\(
\langle\Psi^{+}_{0}|\rho_{\text{CH}}|\Psi^{+}_{0}\rangle 
= 
e^{-2\alpha}/\mathcal{Z}
\), 
\(
\langle\Psi^{-}_{0}|\rho_{\text{CH}}|\Psi^{-}_{0}\rangle 
= 
1/\mathcal{Z}
\) 
and 
\(
\langle\Psi^{+}_{1}|\rho_{\text{CH}}|\Psi^{+}_{1}\rangle 
+
\langle\Psi^{-}_{1}|\rho_{\text{CH}}|\Psi^{-}_{1}\rangle 
=
(1+e^{2\alpha})/\mathcal{Z}
\).
Accordingly, we can't also obtain information on the entanglement of
$\rho_{\text{CH}}$ by the method in Ref.\,\cite{DC}. 

The authors in Ref.\,\cite{YBC} pointed out that the local
operations can decrease the entanglement of the state concerned and an
alternative method is necessary. 
In particular, they discussed a more effective procedure for evaluating
entanglement than the local operations in the D\"ur--Cirac classification in
terms of the mathematical theory of majorization. 
Finally, we comment on a loop--hole in the D\"ur--Cirac classification
from another point view: We attempt to construct a prescription to solve
the problem.  
As has been mentioned, the isotropic state $\rho_{iso}$ is transformed
into $\rho^{\prime}_{iso}$ by the local unitary operator 
$I_{1}\otimes X_{2}$. 
The method in Ref.\,\cite{DC} works for the former but not for
the latter. 
This suggests that, by suitable local unitary operators, the state for
which the D\"ur--Cirac classification doesn't work, could be transformed
into a proper one to which their method is applicable. 
Let us consider the above third example. 
When we use the local unitary operator $I_{1}\otimes X_{2}$, the state
concerned is transformed into 
\(
\rho^{\prime}_{\text{CH}} 
= 
(I_{1}\otimes X_{2})\rho_{\text{CH}}(I_{1}\otimes X_{2})^{\dagger}
\), 
where 
\(
\langle\Psi^{+}_{0}|\rho^{\prime}_{\text{CH}}|\Psi^{+}_{0}\rangle 
= 
1/\mathcal{Z}
\), 
\(
\langle\Psi^{-}_{0}|\rho^{\prime}_{\text{CH}}|\Psi^{-}_{0}\rangle 
= 
e^{2\alpha}/\mathcal{Z}
\), 
and 
\(
\langle\Psi^{+}_{1}|\rho^{\prime}_{\text{CH}}|\Psi^{+}_{1}\rangle 
+
\langle\Psi^{-}_{1}|\rho^{\prime}_{\text{CH}}|\Psi^{-}_{1}\rangle 
=
(e^{-2\alpha}+1)/\mathcal{Z}
\).
Calculating $\Delta$ and $2\lambda_{1}$, we can obtain the
non--trivial expression for $\rho^{\prime}_{\text{CH}}$; for example,
$\rho^{\prime}_{\text{CH}}$ is NPT if and only if $\sinh(2\alpha)>1$. 
If $\rho^{\prime}_{\text{CH}}$ is entangled, the corresponding state
$\rho_{\text{CH}}$ is also considered to be entangled. 
Therefore, we obtain the sufficient condition of the nonseparability for
$\rho_{\text{CH}}$, by the use of the D\"ur--Cirac classification; such a
condition is just $\sinh(2\alpha)>1$. 
This condition is equal to the one which is derived by the use of
Eq.\,(\ref{eq:NPT_CH_wrt_k}). 
It is necessary to examine whether such a prescription is generalized or
not.

\end{document}